\documentclass[preprint,aps,amsmath,amssymb,12pt,showpacs,showkeys]{revtex4} 

\usepackage{bbm}
\usepackage[dvips]{graphicx}
\usepackage{times}
\usepackage{color}
\usepackage{amsmath}
\usepackage{amsfonts,amssymb,graphics,bm,psfrag}

\def\lsim{\mathrel{\rlap{\lower4pt\hbox{\hskip1pt$\sim$}}
    \raise1pt\hbox{$<$}}}                
\def\gsim{\mathrel{\rlap{\lower4pt\hbox{\hskip1pt$\sim$}}
    \raise1pt\hbox{$>$}}}                

\newcommand{\tl}[1]{\widetilde{#1}}

\newcommand{\ul}[1]{\underline{#1}}
\newcommand{\vect}[1]{\mathbf{#1}} 
\newcommand{\mat}[1]{\ul{\rm #1}} 
\newcommand{\h}{\vect{h}}
\newcommand{\n}{\vect{n}}
\newcommand{\R}{\vect{R}}
\newcommand{\x}{\vect{x}}

\newcommand{\hh}{\mat{h}}
\newcommand{\eps}{\varepsilon}

\newcommand{\SO}{{\rm SO}}
\newcommand{\bnabla}{\boldsymbol{\nabla}}
\newcommand{\Id}{\mathbbm{1}}
\newcommand{\mul}{\cdot}
\newcommand{\Dpsi}{D}
\newcommand{\OO}{\mathcal{O}}

\begin{document}

\title{Steering chiral swimmers along noisy helical paths}

\author{Benjamin M. Friedrich}
\email{ben@pks.mpg.de}
\author{Frank J{\"u}licher}
\email{julicher@pks.mpg.de}
\affiliation{
Max Planck Institute for the Physics of Complex Systems\\
{\it N{\"o}thnitzer Stra{\ss}e 38, 01187 Dresden, Germany}}

\date{\today}


\bibliographystyle{apsrev}


\begin{abstract} 
Chemotaxis along helical paths towards a target releasing a chemoattractant
is found in sperm cells and many microorganisms.
We discuss the stochastic differential geometry
of the noisy helical swimming path of a chiral swimmer.
A chiral swimmer equipped with a simple feedback system
can navigate in a concentration gradient of chemoattractant.
We derive an effective equation for the 
alignment of helical paths with a concentration gradient
which is related to the alignment of a dipole in an external field.
We discuss the chemotaxis index in the presence of fluctuations.
\end{abstract}

\maketitle

Biological microswimmers use flagellar propulsion
or undulatory body movements to swim at low Reynolds numbers \cite{Purcell:1977,Lauga:2009}.
In addition to forward propulsion with translational velocity $\vect{v}$,
any chirality in swimming stroke results
in a net angular velocity $\boldsymbol \Omega$.
Hence, such a swimmer moves along a helical path
with curvature $\kappa_0=|{\boldsymbol\Omega}\times\vect{v}|/|\vect{v}|^2$ and
torsion $\tau_0=|{\boldsymbol\Omega}\cdot\vect{v}|/|\vect{v}|^2$
in the absence of fluctuations \cite{Crenshaw:1993a}.
Helical swimming paths have been observed for 
sperm cells 
\nocite{Corkidi:2008}\nocite{Brokaw:1958b}
\cite{Corkidi:2008,Brokaw:1958b,Brokaw:1959,Crenshaw:1996a}, 
eukaryotic flagellates \cite{Jennings:1901,Fenchel:1999}, 
marine zooplankton \cite{McHenry:2003,Jekely:2008},
and even large bacteria \cite{Thar:2001}.
A necessary condition for a pronounced helicity of the swimming path
of a chiral swimmer is given by 
$|\boldsymbol{\Omega}| \gg D_{\rm rot}$
where the rotational diffusion coefficient 
$D_{\rm rot} \sim L^{-3}$
depends strongly on the size $L$ of the swimmer \cite{Berg:2004,Dusenbery:1997}. 
Thus there is a critical size for a chiral swimmer below which 
fluctuations diminish directional persistence and interfere 
with helical swimming.
The bacterium {\it E.~coli} for example is much smaller than the swimmers
mentioned above and fluctuations dominate over an eventual chirality of swimming.
Nevertheless, this bacterium can navigate in a concentration field of a
chemoattractant by performing a biased random walk \cite{Berg:2004}.
A larger swimmer moving along a helical path can exploit a fundamentally 
different chemotaxis strategy:
It has been shown both experimentally 
\cite{Brokaw:1958b,Brokaw:1959,Crenshaw:1996a,Jennings:1901,Fenchel:1999,Thar:2001,Boehmer:2005,Kaupp:2008} 
and theoretically \cite{Crenshaw:1993b,Friedrich:2007}
that such a chiral swimmer can navigate in a concentration gradient of chemoattractant by a simple feedback mechanism.
Here we study the impact of fluctuations and show that sampling a concentration field
along noisy helical paths is a robust strategy for chemotaxis in three
dimensional space even in the presence of noise.
The alignment of noisy helical paths with a concentration gradient is formally equivalent
to the alignment of a polar molecule subject to rotational Brownian motion in an external electrical field.

{\it Stochastic differential geometry of noisy helical paths.}
The geometry of a swimming path $\vect{r}(t)$ is characterized by 
the tangent $\vect{t}=\dot{\vect{r}}/v$, 
normal $\n=\dot {\vect{t}}/\vert \dot {\vect{t}}\vert$ 
and binormal $\vect{b}=\vect{t} \times \n$,
where $v=|\dot{\vect{r}}|$ is speed and
dots denote time derivatives.
The time evolution of these vectors 
can be expressed as \cite{Kamien:2002}
\begin{equation}
\label{frenet3d}
\dot{\vect{r}}=v\,\vect{t},\hspace{2mm} 
\dot{\vect{t}}=v\kappa\,\n,\hspace{2mm}
\dot{\n}=-v\kappa\,\vect{t}+v\tau\,\vect{b},\hspace{2mm}
\dot{\vect{b}}=-v\tau\,\n
\end{equation}
where $\kappa(t)$ and $\tau(t)$
are curvature and torsion of the swimming path $\vect{r}(t)$, respectively.
For a noisy path, $\kappa(t)$ and $\tau(t)$ fluctuate around their mean values
\begin{equation}
\label{curvature_torsion_fluctuations}
\kappa(t)=\kappa_0+\xi_\kappa(t),\quad
\tau(t)=\tau_0+\xi_\tau(t)
\end{equation}
where $\xi_\kappa$ and $\xi_\tau$ are stochastic processes with 
mean zero and respective power spectra
$\tl{S}_{\kappa}$, $\tl{S}_{\tau}$,
as well as a cross power spectrum $\tl{S}_{\kappa,\tau}$
\footnote{Here
$2\pi\tl{S}_{\kappa}(\omega)\delta(\omega-\omega')=\langle \tl{\xi}_\kappa(\omega) \tl{\xi}_\kappa(\omega')^\ast \rangle$
where 
$\tl{\xi}_\kappa(\omega)=
\int_{-\infty}^{\infty}\!\! dt\,\xi_\kappa(t)e^{-i\omega t}$; 
$\tl{S}_{\tau}$ and $\tl{S}_{\kappa,\tau}$ are defined analogously.}.
For simplicity, $v(t)=v_0$ is assumed constant.
The stochastic differential equations (\ref{frenet3d},\ref{curvature_torsion_fluctuations}) 
involve multiplicative noise and should be interpreted in the
Stratonovich sense if $\xi_\kappa$ or $\xi_\tau$ is $\delta$-correlated.

In the noise-free case, $\xi_\kappa=\xi_\tau=0$,
the path $\vect{r}$ is a perfect helix
with radius 
$r_0=\kappa_0/(\kappa_0^2+\tau_0^2)$,
pitch
$2\pi h_0= 2\pi\,\tau_0/(\kappa_0^2+\tau_0^2)$,
and helix angle
$\theta_0=\tan^{-1}(h_0/r_0)$.
We define the helix reference frame $(\R,\h_1,\h_2,\h_3)$ 
by the linear transformation
\begin{equation}
\label{helix_frame}
\R=\vect{r}+r_0 \n,\quad
\h_3=\sin\theta_0\,\vect{t}+\cos\theta_0\,\vect{b},
\end{equation}
$\h_1=-\n$ and $\h_2=\h_3\times\h_1$.
Here, $\R(t)$ is the centerline of the helical path $\vect{r}(t)$ 
and $\h_3$ is called the helix vector.
The helix frame can be interpreted
as the material frame of a solid disk
with center $\R$, see Fig.~1A:
The disk translates and rotates
such that a marker point on the disk's circumference
traces the helical path $\vect{r}$.
For a perfect helix,  
$\dot{\R}=\overline{v}\h_3$,
$\dot{\h}_3=0$, 
$\dot{\h}_1=\omega_0 \h_2$,
$\dot{\h}_2=-\omega_0 \h_1$
where
$\overline{v}=\omega_0 h_0$
and
$\omega_0=v_0\, (\kappa_0^2+\tau_0^2)^{1/2}$
is the frequency of helical swimming.
The period of a helical turn is $T=2\pi/\omega_0$.

In the presence of fluctuations, 
$\dot{\R}=\overline{v}\h_3+v_0 r_0 (\xi_\tau\vect{b}-\xi_\kappa\vect{t})$.
The helix vector $\h_3$ performs a stochastic motion on the unit sphere
which is characterized by
\begin{equation}
\langle \h_3(0)\cdot\h_3(t)\rangle \approx \exp(-t/t_P)
\end{equation}
for times $t$ longer than the correlation time of curvature and torsion fluctuations.
In the following, we determine the persistence time $t_P$ in a limit of weak noise;
the result is given in eqn.~(\ref{tP}). 
The rotation matrix
$\mat{H}(t)$ with ${H_{kl}=\h_k(0)\cdot\h_l(t)}$
is an element of $\SO(3)$.
The Lie algebra of $\SO(3)$ is spanned by the infinitesimal
rotations $\mat{E}_j$ with $(\mat{E}_j)_{kl}=\epsilon_{kjl}$, $j=1,2,3$.
The time evolution of $\mat{H}(t)$ is given by a matrix-valued differential
equation
$\dot{\mat{H}}=\mat{H}\mul\hh$
with infinitesimal rotation 
$\hh=\omega_0\mat{E}_3+\xi_j\mat{E}_j$ 
where we use Einstein summation convention for $j=1,2,3$. 
From eqns.\ (\ref{frenet3d}-\ref{helix_frame}), we find
$\xi_1=0$,
$\xi_2=\omega_0\,(r_0\xi_\tau-h_0\xi_\kappa)$, and 
$\xi_3=\omega_0\,(r_0\xi_\kappa+h_0\xi_\tau)$.
The rotation of the helix frame after a time $t$ consists of
a rotation around $\h_3(0)$ by an angle $\omega_0 t$
and random rotations around all axes due to the curvature and torsion fluctuations.
We characterize these random rotations by 
continuous stochastic processes $\Xi_j(t)$ with $\Xi_j(0)=0$ and write
\begin{equation}
\label{HnT}
\mat{H}(t)=\exp(\omega_0 t\,\mat{E}_3)\cdot\exp( \Xi_j \mat{E}_j ).
\end{equation}
Note that $\exp(\omega_0 nT\,\mat{E}_3)=\Id$ after $n$ helical turns.
The $\Xi_j$ represent generalized rotation angles:
$\Xi_1$ and $\Xi_2$ describe rotations of $\h_3$.
Symmetry implies 
$\langle\Xi_1\rangle=\langle\Xi_2\rangle=0$.
We consider the limit of weak noise characterized by 
$|\tl{S}_\kappa(\omega)|,|\tl{S}_\tau(\omega)|\ll \kappa_0/v_0$.
One can develop a systematic expansion in powers of the noise strengths 
$\eta_\kappa$, $\eta_\tau$
with 
$\eta_\kappa^2=v_0 r_0 \int_{-\infty}^{\infty} dt |\langle \xi_\kappa(0)\xi_\kappa(t) \rangle|$
and analogously $\eta_\tau$.
We write $\cong$ to denote equality to leading order in 
$\eta_\kappa$, $\eta_\tau$.
For times $t=nT$ longer than the correlation time of $\xi_\kappa$, $\xi_\tau$
but still shorter than $T/(\eta_\kappa+\eta_\tau)$,
we find
\begin{equation}
\label{Xi2}
\langle \Xi_1^2\rangle \cong \langle \Xi_2^2\rangle \cong 2D\,t,\quad
\langle\Xi_1\,\Xi_2\rangle \cong 0
\end{equation}
with
$4D=\tl{S}_2(\omega_0)$
where 
$\tl{S}_2(\omega)=\omega_0^2[h_0^2 \tl{S}_{\kappa}(\omega)
+r_0^2 \tl{S}_{\tau}(\omega)
-2r_0 h_0{\rm Re\,} \tl{S}_{\kappa,\tau}(\omega)]$
is the power spectrum of $\xi_2$.
Hence, the stochastic motion of the helix vector $\h_3$ 
can be effectively described as isotropic rotational diffusion
with rotational diffusion coefficient $D$ for long times.
The derivation of (\ref{Xi2}) proceeds as follows:
$\mat{H}(t)$ can be written as a time-ordered 
exponential integral $\mat{H}(t) = {\rm {\bf T}exp} \int_0^{t} dt'\, \hh(t')$.
To linear order in the noise strengths,
${\Xi_2+i\,\Xi_1} \cong \int_0^{t}\hspace{0mm} dt'\, \xi_2(t)\, e^{i\omega_0 (t-t') }$.
Next,
$\langle\Xi_1^2+\Xi_2^2\rangle 
\cong\int_0^{t}\hspace{0mm} dt_1\int_0^{t}\hspace{0mm} dt_2\, 
\langle\xi_2(t_1)\,\xi_2(t_2)\rangle\,e^{-i\omega_0(t_1-t_2)}
\approx \,\tl{S}_2(\omega_0)\,t$. 
Similarly, $\langle\Xi_3^2\rangle\cong \tl{S}_3(0)\,t$
where $\tl{S}_3(\omega)$ is the power spectrum of $\xi_3$.
%
Hence the swimming path $\vect{r}(t)$ is a noisy helix with a centerline $\R(t)$ 
that follows a persistent random walk 
(on time-scales larger than the
correlation time of curvature and torsion fluctuations) 
\endnote{
For a related result concerning DNA, see:
N.~B.~Becker and R.~Everaers, Phys. Rev. E. {\bf 76}, 021923 (2007).}%
.
This persistent random walk 
has a persistence time 
\begin{equation}
\label{tP}
t_P=(2D)^{-1}=2/\tl{S}_2(\omega_0)
\end{equation}
that is governed by the power spectra of the curvature and torsion
fluctuations evalutuated at the helix frequency $\omega_0$
and a persistence length
$l_P=\overline{v}\,t_P$
\cite{Kamien:2002,Friedrich:2008a}.

{\it A chemotactic chiral swimmer.}
We now consider a chiral swimmer in a concentration field $c(\x)$ of chemoattractant
equipped with a feedback mechanism which allows it to dynamically adjust its curvature 
and torsion in response to a chemotactic stimulus $s(t)$.
The stimulus $s(t)=\sum_j \delta(t-t_j)$ counts single chemoattractant molecules detected by the swimmer at times $t_j$.
The rate $q=\langle s\rangle$ of molecule detection 
of the swimmer is assumed 
proportional to the local chemoattractant concentration \cite{Friedrich:2008b}
\begin{equation}
\label{stimulus}
\langle s(t) \rangle = q(t) = \lambda\,c(\vect{r}(t)).
\end{equation}
When $q$ is large compared to a typical measurement time $\sigma^{-1}$ of the swimmer and
$q(t)$ changes on a time-scale slow compared to the mean inter-event-interval $1/q$,
then we can replace $s(t)$ by a coarse-grained version known as the diffusion limit
\begin{equation}
\label{diffusion_limit}
s(t)\approx q(t)+\sqrt{q(t)}\,\xi_s(t)
\end{equation}
where $\xi_s(t)$ is Gaussian white noise with 
$\langle \xi_s(t_1)\xi_s(t_2)\rangle=\delta(t_1-t_2)$.
In this limit $\eta\ll 1$ where $\eta=(q \sigma)^{-1/2}$
characterizes the relative noise strength of $s(t)$ 
for an averaging time $\sigma$ \cite{Friedrich:2008b}.
The chemotactic stimulus $s(t)$ is transduced by a signaling system of the swimmer and 
triggers a chemotactic response which we characterize by a dimensionless
output variable $a(t)$ with $a=1$ for a time-independent stimulus $s(t)=s_0$.
We assume that $a(t)$ affects curvature and torsion in a linear way
$\kappa(t) = \kappa_0+\kappa_1 (a(t)-1)$ and analogously for $\tau$ \cite{Friedrich:2007}.
Recall that swimming speed $v(t)=v_0$ is assumed constant.
For the signaling system relating stimulus $s(t)$ and output $a(t)$, 
we use a simple dynamical system 
which exhibits adaptation and a relaxation dynamics
\nocite{barkai_leibler97}
\cite{barkai_leibler97,Friedrich:2007,Friedrich:2008b}
\begin{equation}
\label{signaling}
\begin{split}
\sigma\, \dot{a} = p\, s - a, \quad
\mu\, \dot{p} &= p\, (1 - a). \\
\end{split}
\end{equation}
Here $p(t)$ is a variable representing a dynamic sensitivity;
$\sigma$ is a relaxation time and $\mu$ is a time-scale of adaptation.
For a time-independent stimulus $s(t)=s_0$, the system
(\ref{signaling}) reaches a stationary state
with $a=1$, $p=1/s_0$.
Small periodic variations of the stimulus 
$s(t)=s_0+s_1 \cos\omega t$
evoke a periodic response of the output variable
$a(t)=1+s_1{\rm Re\,} \tl{\chi}_a(\omega)e^{i\omega t}+\OO(s_1^2)$
with linear response coefficient
$\tl{\chi}_a(\omega)=i\omega\mu/[s_0(1+i\omega\mu-\sigma\mu\omega^2)]$.

{\it Swimming in a concentration gradient.}
We consider a chemotactic chiral swimmer in a linear
concentration field of chemoattractant 
\begin{equation}
\label{linear_concentration_field}
c(\x)=c_0+\vect{c}_1 \cdot\x.
\end{equation}
Fig.~1B shows an example of a stochastic swimming path $\vect{r}(t)$ 
in such a linear concentration field
which has been obtained numerically.
In the simulation, the chemotactic chiral swimmer detects individual
chemoattractant moleculs arriving at random times 
(distributed according to an inhomogenous Poisson process with rate $q(t)$ \cite{Friedrich:2008b}). 

We characterize the chemotaxis mechanism of a chiral swimmer 
in the limit where both
chemoattractant concentration $c_0$ is high with 
$\eta=(\lambda c_0\sigma)^{-1/2}\ll 1$, and
the concentration gradient is weak with $\nu=|\vect{c}_1| r_0/c_0\ll 1$.
The concentration gradient $\vect{c}_1$ is a sum 
$\vect{c}_1=c_\parallel\h_3+\vect{c}_\perp$ with
a component 
parallel to $\h_3$ 
of length $c_\parallel=\vect{c}_1\cdot\h_3$, and 
a component $\vect{c}_\perp=\vect{c}_1-c_\parallel\h_3$ perpendicular to $\h_3$
of length $c_\perp=|\vect{c}_\perp|$.
While the swimmer moves in the concentration field along the noisy helical
path $\vect{r}(t)$, the binding rate $q(t)$ varies with time.
In the limit of weak noise and a weak gradient,
we approximate $q(t)$ by the value for $q(t)$ 
obtained for swimming along the unperturbed path with
chemotactic feedback switched off
$q(t)\lambda \approx c_0 + c_\parallel(0) \overline{v} t + c_\perp(0) r_0 \cos(\omega_0 t+\varphi_0)$
where $\varphi_0$ is the angle enclosed by $\vect{c}_\perp(0)$ and $\h_1(0)$.
It is this periodic modulation of $q(t)$ which underlies navigation
in a concentration gradient as it causes a bias
in the orientational fluctuations of $\h_3$:
The stimulus $s(t)$ elicits a periodic modulation of the average curvature
and torsion with amplitude proportional to $c_\perp$.
As a consequence, the expectation values 
$\langle\Xi_1\rangle$, $\langle\Xi_2\rangle$
of the generalized rotation angles introduced in (\ref{HnT})
are non-zero and scale with $c_\perp$,
$\langle \Xi_2+i\Xi_1 \rangle \cong 
c_\perp\,\eps\tl{\chi}_a(\omega_0)e^{-i\varphi_0}\,t$
with
$\eps=\lambda \omega_0 r_0 (h_0\kappa_1-r_0\tau_1) / 2$.
Similarly,
$\langle \Xi_3 \rangle \cong c_\parallel\,\overline{\eps}\,t$
with $\overline{\eps}=\mu \omega_0^2 h_0 ( r_0 \kappa_1 + h_0 \tau_1 )/c_0$ 
\cite{Friedrich:2007}.

We can now derive an effective stochastic equation of motion for the helix frame
in the limit $\eta,\nu\ll 1$
by a coarse-graining procedure as outlined in \cite{Friedrich:2008b}.
The Stratonovich stochastic differential equation for the helix frame
\begin{equation}
\label{motion3d_full_noise}
\begin{array}{lll}
\dot{\R} &= \overline{v}\, \h_3 & \\
\dot{\h}_3 &= - \eps\, {\rm Re\,}\hspace{-1mm}\left[ \tl{\chi}_a(\omega_0)\,\vect{c} \right]
& +\, \overline{\xi}_1\h_2-\overline{\xi}_2\h_1 \\
\dot{\h}_j &= 
-(\dot{\h}_3\cdot\h_j)\,\h_3 
&+\, \epsilon_{kj3}\,\overline{\omega}\,\h_k,\quad j=1,2 \\
\end{array}
\end{equation}
generates the statistics of the noisy helical path
to leading order in $\nu$ and $\eta$
\footnote{Eqn.~(\ref{motion3d_full_noise}) generates stochastic 
processes as defined in (\ref{HnT}) whose second-order statistics
is correct to leading order in $\nu$ and $\eta$%
.}.
Here 
$\overline{\omega}=\omega_0+\overline{\eps}c_\parallel$  and $\vect{c} = \vect{c}_\perp + i\, \h_3 \times \vect{c}_\perp$.
Eqn.~(\ref{motion3d_full_noise}) 
contains a multiplicative noise term $\overline{\xi}_1\h_2-\overline{\xi}_2\h_1$
where $\overline{\xi}_j$ denotes Gaussian white noise with
$\langle \overline{\xi}_k(t_1)\,\overline{\xi}_l(t_2)\rangle=2D\,\delta_{kl} \delta(t_1-t_2)$.
Here $D$ plays the role of a rotational diffusion coefficient and is given by
$D=|\eps\,\tl{\chi}_a(\omega_0)/r_0|^2\,c_0/\lambda$.
Note that $D$ is concentration dependent with $D\sim 1/c_0$.
In the deterministic limit $\overline{\xi}_1=\overline{\xi}_2=0$, we recover the results
from \cite{Friedrich:2007}.
Eqn.~(\ref{motion3d_full_noise}) 
provides a coarse-grained description of 
the time evolution of the helix frame on time-scales
larger than the correlation time $\sigma$ of curvature and torsion fluctuations. 

{\it Effective dynamics of the alignment angle.}
In a linear concentration field (\ref{linear_concentration_field}),
the quantity of interest is the alignment angle $\psi$ 
enclosed by the helix vector $\h_3$ and the direction of the
gradient $\vect{c}_1$ \cite{Friedrich:2007}, see Fig.~1A.
The symmetries of the problem imply that the dynamics of $\psi$ 
decouples from the other degrees of freedom of the helix frame. 
From 
(\ref{motion3d_full_noise}), we find by using the rules of stochastic calculus
\begin{equation}
\label{motion3d_linear_noise}
\dot{\psi} = - \beta \sin\psi + \xi + \Dpsi \cot\psi.
\end{equation}
Here $\xi$ denotes Gaussian white noise with 
$\langle \xi(t_1)\,\xi(t_2)\rangle=2\Dpsi\,\delta(t_1-t_2)$.
The alignment rate is $\beta=-|\bnabla c|\, \eps\, {\rm Re\,}\tl{\chi}_a(\omega_0)$.
In the absence of fluctuations, we recover the
deterministic limit $\dot{\psi}=-\beta\sin\psi$
\cite{Friedrich:2007}.
In this limit, the steady state is characterized by either
parallel alignment of helix vector and concentration gradient with $\psi=0$ for $\beta>0$ 
or by anti-parallel alignment with $\psi=\pi$ for $\beta<0$.
Eqn.~(\ref{motion3d_linear_noise}) contains a noise-induced drift term
$D\cot\psi$ 
which diverges for $\psi=0$ and $\psi=\pi$ 
implying that noise impedes perfect parallel or anti-parallel alignment 
of the helix vector.

The corresponding Fokker-Planck equation for the probability distribution $P(z,t)$ of
$z=\cos\psi$
with 
$|z|\le 1$
reads
$\dot{P}=-\partial_z[{(1-z^2)}{(\beta-\Dpsi\partial_z)}]P$.
Fig.~1C compares $P(z,t)$ to a histogram of $z$ obtained
from simulating $10^5$ chemotactic chiral swimmers in a linear concentration
field. 
The distribution $P(z,t)$ relaxes to a steady state distribution 
$P_0(z)\sim \exp(\beta z/\Dpsi)$
on a time-scale which is set by the inverse alignment rate $\beta^{-1}$.
%
This steady-state distribution $P_0(z)$
has its maximum at $z^{\ast}=\pm 1$
for $\beta \gtrless 0$, respectively.
The first moment of $P_0(z)$ is given by the Langevin function \cite{Debye:1929}
\begin{equation}
\label{zmean}
\langle z \rangle = {\rm coth}({\rm Pe})-{\rm Pe}^{-1},\quad
{\rm Pe}=\beta/\Dpsi
\end{equation}
where ${\rm Pe}$ describes a Peclet number of rotational motion.
Note that this result for the mean orientation of a chemotactic
chiral swimmer is formally equivalent to the 
orientation of a polar molecule in an external electrical field:
Eqn.\ (\ref{zmean}) with 
${\rm Pe}$ replaced by $|\vect{m}||\vect{E}|/(k_BT)$
also describes the mean orientation 
$\langle z\rangle=\langle\vect{m}\cdot\vect{E}\rangle/(|\vect{m}||\vect{E}|)$
of a polar molecule with dipole moment $\vect{m}$
subject to rotational Brownian motion
in an electric field $\vect{E}$ \cite{Debye:1929}.
Note that eqn.~(\ref{zmean}) characterizes an 
active process while a polar molecule 
is an equilibrium system.

\begin{figure}[tb]
\includegraphics[width=8.5cm]{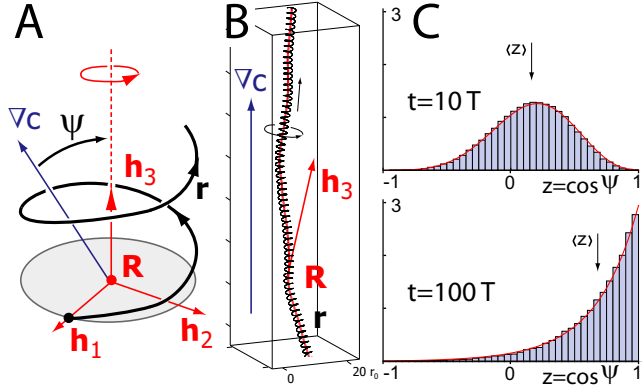}
\caption{
{\bf A.} A helical path $\vect{r}$ can be described as the trajectory of a point on
the circumference of a disk which rotates and translates; see text for details.
{\bf B.} Helical swimming path $\vect{r}$ of a chemotactic chiral swimmer
in a linear concentration field: The helix vector $\h_3$ fluctuates 
around the direction of the concentration
gradient $\bnabla c$.
Parameters were $v_0=r_0/\sigma$, $\tau_0=0.2/r_0$,
$-\kappa_1=\tau_1=0.5/r_0$, $\mu=\sigma$,
$\lambda c_0=10^2/\sigma$, $|\vect{c}_1|=10^{-2}c_0/r_0$.
{\bf C.} Histograms of $z=\cos\psi$ where $\psi$ equals the angle enclosed by
the helix axis and the gradient direction for a simulated ensemble of helical
swimming paths as in B with initial distribution $P(z,0)=\delta(z)$
at times $t=10T,100T$. 
Also shown is the analytical solution $P(z,t)$ (red).
Approximately, $P(z,100T)$ equals the steady-state distribution
$P_0(z)\sim\exp({\rm Pe}\,z)$.
} 
\end{figure}

At steady state, 
a chemotactic chiral swimmer
moves up a concentration gradient with average speed
$\langle z\rangle\overline{v}$.
The chemotaxis index $\rm CI$ is 
defined as the ratio
of this average speed gradient-upwards and the swimming speed $v_0$
\begin{equation}
{\rm CI}=\langle z\rangle{\rm CI_{max}},\quad
{\rm CI_{max}}=\overline{v}/v_0=\sin\theta_0.
\end{equation}
Note that ${\rm CI}$ approaches its maximal value $\rm CI_{max}$ 
for ${\rm Pe}\gsim 1$.
This condition is satisfied already beyond moderate concentration gradients
with $|\bnabla c|\gsim |\eps|/(\lambda r_0)^2$.
The maximal value $\rm CI_{max}$ for the chemotaxis index is limited only by the geometry of
helical swimming.

{\it Relation to experiments.}
Chemotaxis of sperm cells has been extensively studied for sea urchin sperm
cells \cite{Kaupp:2008}.
Tracking experiments in three dimensions 
show that these sperm cells swim along noisy helical paths 
with typical values for swimming speed, average curvature and torsion
$v_0\approx 100-200\,\mu{\rm m}\,{\rm s}^{-1}$, 
$\kappa_0\approx 0.025-0.05\,\mu{\rm m}^{-1}$, 
$\tau_0\approx -0.0025\,\mu{\rm m}^{-1}$ \cite{Crenshaw:1996a,Corkidi:2008}.
For comparision, the length of the sperm tail is ${L\approx 50\,\mu{\rm m}}$ \cite{Boehmer:2005}. 
Using a two-dimensional experimental setup in which sperm cells swim
along a circular path, it has been shown that a periodic chemotactic stimulus
causes a phase-locked periodic swimming response \cite{Boehmer:2005,Wood:2005}.
Such a behavioural response is consistent with our model of a chemotactic chiral swimmer.

In a pioneering experiment, C.~J.~Brokaw observed
helical swimming paths of bracken fern sperm cells 
in a shallow observation chamber 
\cite{Brokaw:1958b,Brokaw:1959}
\footnote{Bracken fern sperm cells are very different from animal sperm
cells, but also motile with thrust generated by several flagella.}. 
In the absence of chemoattractant, 
sperm swimming paths were noisy helices whose centerlines
could be described as planar persistent random walks with 
persistence time $t_{P,\rm 2d}\approx 5\,{\rm s}$ and
net speed $\overline{v}\approx 200\,{\mu\rm m}\,{\rm s}^{-1}$,
corresponding to a persistence length of $l_{P,\rm 2d}\approx 1\,{\rm mm}$.
Accordingly, the planar orientational fluctuations of the helix vector are
characterized by a rotational diffusion coefficient $D=t_{P,\rm 2d}^{-1}=0.2\,{\rm s}^{-1}$.
In a strong concentration gradient of chemoattractant,
sperm swimming paths were bent helices which aligned with
the gradient direction at a rate proportional to the relative
strength of the concentration gradient 
$\beta\approx 150\,{\mu \rm m}\,{\rm s}^{-1}\, |\bnabla c|/c$
\footnote{The value quoted in \cite{Friedrich:2007} was incorrect
due to a confusion of decadic and natural logarithm.}. 
In an initially homogeneous concentration field of charged chemoattractant,
alignment of helical sperm swimming paths 
could also induced by applying an external electrical field $|\vect{E}|$.
In this case, it was found that the alignment
rate is proportional to the field strength
$\beta/|\vect{E}|\approx 1.6\,{\rm s}^{-1}\,{\rm (V/cm)}^{-1}$.
The mean alignment $\langle z\rangle=\langle \h_3\cdot \vect{E}\rangle/|\vect{E}|$ 
of helical paths at steady state was measured as a function of field strength $|\vect{E}|$.
The experimental data could be well fitted by eqn.\ (\ref{zmean})
assuming ${\rm Pe}\sim |\vect{E}|$ and yielded 
${\rm Pe}/|\vect{E}|\approx 8\,{(\rm V/cm)}^{-1}$
\endnote{Swimming of sperm cells was restricted 
to a shallow observation chamber 
of height $100\,{\mu\rm m}$ 
which affects the statistics of $z$:
If the helix vector is constrained to a plane parallel to $\vect{E}$, 
our theory predicts $\langle z\rangle_{\rm 2d}=I_1({\rm Pe})/I_0({\rm Pe})$.
This also describes the data in \cite{Brokaw:1959}
and gives ${\rm Pe}/|\vect{E}|\approx 5\,{\rm (V/cm)}^{-1}$.}.
The above estimates for $D$ and $\beta/|\vect{E}|$ give
approximately the same value for 
${\rm Pe}/|\vect{E}|=(\beta/D)/|\vect{E}|$ \cite{Brokaw:1959}.
The physical origin of helix alignment in an electrical field is not entirely known:
The observed alignment might be due to electrohydrodynamic effects resulting
from sperm cells binding chemoattractant ions 
(with sperm cells effectively behaving as electric dipoles) \cite{Brokaw:1959}.
An alternative possibility is that the electric field induces a concentration gradient of chemoattractant ions and that the observed alignment of helical paths 
is a result of chemotactic navigation in this gradient.

{\it Conclusion.}
In this Letter, we studied 
the stochastic differential geometry of noisy helical swimming paths
which is relevant for many biological mircoswimmers with chiral propulsion 
\cite{Corkidi:2008,Brokaw:1958b,Brokaw:1959,Crenshaw:1996a,Jennings:1901,Fenchel:1999,McHenry:2003,Jekely:2008,Thar:2001}.
A simple feedback mechanism enables a chiral swimmer to navigate 
along a helical path upwards a concentration gradient of chemoattractant.
Chemotaxis along noisy helices is employed by sperm cells and possbily other
biological microswimmers 
\cite{Brokaw:1958b,Brokaw:1959,Crenshaw:1996a,Jennings:1901,Fenchel:1999,Thar:2001}.
A similar mechanism underlies phototaxis of the
unicellular flagellate {\it Chlamydomonas} \cite{Crenshaw:1996b},
and is found in phototactic marine zooplankton
\cite{McHenry:2003,Jekely:2008}.
Our theory shows that navigation along helical paths is remarkably robust in the presence of fluctuations:
An effective rotation of the helix vector is determined by 
integrating its orientational fluctuations over several helical turns.
Consequently, a small bias in these orientational fluctuations due to
chemotactic signaling results in robust steering
and the helix vector tends to align with the concentration gradient $\bnabla c$.
If chemotactic signaling is adaptive, the alignment rate $\beta$ is
proportional to the relative strength of the
concentration gradient $|\bnabla c|/c$.
After a transient period of alignment of duration $\beta^{-1}$, a chemotactic chiral swimmer
moves upwards the concentration gradient with an average speed
that is only limited by the geometry of helical swimming provided the strength of the concentration
gradient exceeds a characteristic value.
%
%
We conclude that temporal sampling of a concentration field along
a helical path provides a universal strategy for chemotaxis
which is highly adapted for a noisy environment.


\end{document}